\documentclass[a4paper,12pt]{article}
\usepackage[utf8x]{inputenc}

\title{Drude and Superfluid Weights in Extended Systems: the Role of
  Discontinuities and $\delta$-peaks in the One and Two-Body Momentum
  Densities}
\author{Bal\'azs Het\'enyi
\footnote{Present address: Department of Physics,
Bilkent University, 06800 Bilkent, Ankara, Turkey} \\ \\
{\it Institute for Theoretical Physics} \\
{\it Graz University of Technology} \\
{\it A-8010 Graz, Austria} \\
\\ \\
{\it Max Planck Institute for the Physics of Complex Systems}\\
{\it 01187 Dresden, Germany} \\ \\
{\it Department of Physics}\\
{\it Bilkent University}\\
{\it 06800 Bilkent, Ankara, Turkey} \\
}

\begin{document}

\maketitle

\begin{abstract}
 The question of conductivity is revisited.  Using the total
  momentum shift operator to construct the perturbed many-body
  Hamiltonian and ground state wave function the second derivative of
  the ground state energy with respect to the perturbing field is
  expressed in terms of the one and two-body momentum densities.  The
  distinction between the adiabatic and envelope function derivatives,
  hence that between the Drude and superfluid weights can be
  introduced in a straightforward manner.  It is shown that a
  discontinuity in the momentum density leads to a contribution to the
  Drude weight, but not the superfluid weight, however a
  $\delta$-function contribution in the two-body momentum density
  (such as in the BCS wave-funtion) contributes to both quantities.
  The connection between the discontinuity in the momentum density and
  localization is also demonstrated.
\end{abstract}

To distinguish between conductors and insulators an expression for the
frequency-dependent conductivity was derived by Kohn~\cite{Kohn64}.
The DC conductivity (Drude weight) corresponds to the strength of the
$\delta$-function peak of the conductivity at zero frequency.  The
Drude weight is often expressed~\cite{Kohn64,Shastry90} in terms of
the second derivative of the ground state energy with respect to a
phase associated with the perturbing field.  This phase has the effect
of shifting the momenta of the system.  Scalapino, White, and Zhang
(SWZ)~\cite{Scalapino92,Scalapino93} have pointed out that taking the
derivative with respect to the phase is ambiguous: if the derivative
is defined via adiabatically shifting the state which is the ground
state at zero field, then the Drude weight results.  In the presence
of level crossings the adiabatically shifted state may be an excited
state for finite perturbation.  The superfluid weight is obtained if
the derivative corresponds to the ``envelope function'', i.e. the
ground state for any value of the perturbation.  SWZ also state that
nonadiabatic crossings occur infinitesimally close to zero field if
the dimensionality is greater than one.

In this paper this question is revisited.  Based on the total momentum
shift operator~\cite{Hetenyi09} the perturbed Hamiltonian and ground
state wavefunction are explicitly constructed.  This operator plays an
important role in constructing the total position operator for
many-body systems~\cite{Hetenyi09,Resta98,Resta99}.  The second
derivative of the ground state energy with respect to the perturbing
field is then expressed in terms of the one and two-body momentum
densities.  It is then shown that the adiabatic and envelope
derivatives can be distinguished by varying the length scale
associated with the total momentum shift operator, which is also the
length scale of the perturbing field.  When this length scale is
assumed to be the same as the size of the system then
$\frac{\partial^2 E(\Phi)}{\partial \Phi^2}$ is proportional to the
superfluid weight, if this length scale is assumed to be much larger
than the system size than $\frac{\partial^2 E(\Phi)}{\partial \Phi^2}$
corresponds to the Drude weight.  For continuous one and two-body
momentum densities both quantities are zero.  If the one-body momentum
density is discontinuous then the Drude weight is finite, but the
superfluid weight is zero, and if the two-body momentum displays a
$\delta$-peak (Cooper pairing) then both the Drude and superfluid
weights are finite.  Hence insulators, metals, and superconductors can
be distinguished.  While a discontinuous momentum density being a sign
of conduction is a well-known result of many-body
theory~\cite{Mahan00} and plays an important role in the Landau theory
of Fermi liquids~\cite{Mahan00,Negele98}, the foundations of the
latter are distinct from those for the conductivity put forth by
Kohn~\cite{Kohn64}.  In this work the finiteness of the Drude weight
and the discontinuity in the momentum density are shown to coincide.
Moreover, it is also demonstrated that the localization tenet
suggested by Kohn~\cite{Kohn64}, namely that a system 
localized(delocalized) in the many-body configuration space is
insulating(metallic), is also equivalent to the absence(presence) of a
discontinuity in the momentum density.  Hence the Landau theory of
Fermi liquids and the localization theory of Kohn are placed on the
same theoretical footing.

We consider a system of interacting fermions whose Hamiltonian is
periodic in $L$.  We will assume that the ground state is also
periodic in $L$ (i.e. $\Phi=0$).  This leads to no loss of generality,
since if the ground state is at a finite $\Phi$, the Hamiltonian can
be shifted.  We wish to write the Hamiltonian for such a system.  We
first write
\begin{equation}
\hat{\mathcal{H}} = \mathcal{H}(\{g(k)\};\{\hat{c}_k^{(\dagger)}\})
\end{equation}
where $g(k)$ are continuous functions of $k$ and
$\hat{c}_k^{(\dagger)}$ denote creation and annihilation operators of
particles at wave-vector $k$.  This Hamiltonian includes only states
which are periodic in $L$.  Due to the periodicity the spacing of the
points on which the momenta are represented is $\Delta k = 2\pi /L$.
$\hat{\mathcal{H}}$ is not the full Hamiltonian of the system, since
the states with twisted boundary conditions (which correspond to
$k$-vectors which fall between the grid-points) do not appear as
eigenstates.  To include them we write
\begin{equation}
  \hat{\mathcal{H}}(\alpha) =
\mathcal{H}(\{g(k+\alpha)\};\{\hat{c}_{k+\alpha}^{(\dagger)}\}).
\end{equation}
Here all the $k$ vectors have been shifted by $\alpha$, however, the
spacing of the $k$-vectors is unchanged.  The full Hamiltonian can be
written
\begin{equation}
\hat{\mathcal{H}}_T = \frac{1}{2\pi}\int_{-\pi}^{\pi} \mbox{d}\alpha
\hat{\mathcal{H}}(\alpha).
\end{equation}
This Hamiltonian is the full Hamiltonian in the sense that the system
itself is periodic in $L$, however states of all boundary twists are
included.  $\hat{\mathcal{H}}_T$ is block-diagonal, since Hamiltonians
with different values of $\alpha$ correspond to different Hilbert
spaces.  In the limit $L\rightarrow\infty$ $\hat{\mathcal{H}}_T$
becomes the full Hamiltonian of the infinite system.

To stress this point one can consider the Hubbard model for a system
with size $L$ with Hamiltonian written in reciprocal space,
\begin{equation}
  \hat{H}_{Hub} = \sum_{k \sigma} \epsilon_{k\sigma} n_{k\sigma}+ 
U \sum_{k k' q} \hat{c}_{k\uparrow}^\dagger \hat{c}_{k'\downarrow}^\dagger
\hat{c}_{k+q\uparrow} \hat{c}_{k'-q\downarrow}.
\end{equation}
The eigenstates of this Hamiltonian are periodic in $L$.  The spacing
of the $k$-vectors is $\Delta k = 2\pi/L$.  
The shifted Hubbard Hamiltonian 
\begin{equation}
  \hat{H}_{Hub} = \sum_{k \sigma} \epsilon_{k+\alpha\sigma} n_{k+\alpha\sigma}+ 
U \sum_{k k' q} \hat{c}_{k+\alpha\uparrow}^\dagger
\hat{c}_{k'+\alpha\downarrow}^\dagger
\hat{c}_{k+q+\alpha\uparrow} \hat{c}_{k'-q+\alpha\downarrow},
\end{equation}
has eigenstates with twisted boundary conditions, however, the
Hamiltonian still corresponds to a system periodic in $L$, as the
spacing between the $k$-vectors is still $\Delta k = 2\pi/L$.

It is expedient to introduce the total momentum shift operator
\begin{equation}
\hat{U}\left(\frac{2\pi}{L}\right) = \mbox{exp}\left(i\frac{2\pi
\hat{X}}{L}\right),
\end{equation}
where $\hat{X}=\sum_ii\hat{n}_i$, the sum of the positions of all the particles,
and which has the property that~\cite{Hetenyi09}
\begin{equation}
\hat{U}\left(\frac{2\pi}{L}\right)  \hat{c}_k = \left\{ \begin{array}{rl}
  \hat{c}_{k - \frac{2\pi}{L}} \hat{U}, & k = 2\frac{2\pi}{L},...,2\pi \\
  \hat{c}_{2\pi} \hat{U},   & k = \frac{2\pi}{L}.
       \end{array} \right.
\end{equation}
We extend $\hat{U}(2\pi/L)$ to lengths $nL$ with $n$ integer.  Then
momentum shifts to states with twisted boundary conditions on $L$ are
also included.  Taking the limit $n\rightarrow\infty$ we can write
\begin{equation}
\hat{U}(\gamma) \hat{\mathcal{H}}(\alpha)\hat{U}(-\gamma)
=  \mathcal{H}(\{g(k+\alpha)\};\{\hat{c}_{k+\alpha-\gamma}^{(\dagger)}\}),
\end{equation}
for arbitrary $\gamma$ thus
\begin{equation}
\label{eqn:Htg}
\hat{U}(\gamma) \hat{\mathcal{H}}_T \hat{U}(-\gamma)
=  \frac{1}{2\pi}\int_{-\pi}^{\pi} \mbox{d}\alpha
\mathcal{H}(\{g(k+\alpha+\gamma)\};\{\hat{c}_{k+\alpha}^{(\dagger)}\}).
\end{equation}
The transformed Hamiltonian defined in Eq. (\ref{eqn:Htg}) has the
same eigensystem as $\hat{\mathcal{H}}_T$.  The transformation merely shifts the
block diagonal Hamiltonians which comprise $\hat{\mathcal{H}}_T$.

The linear response of a system with periodic boundary conditions can
be cast using the total momentum shift.  We assume that the system of
interest has a Hamiltonian of the form
\begin{equation}
\hat{H} = \sum_k \epsilon_k \hat{n}_k + \hat{H}_i,
\end{equation}
where $\hat{H}_i$ denotes an interaction diagonal in the coordinate
representation.  This Hamiltonian includes the ground state, which is
also periodic in $L$.  For the ground state wavefunction we assume the
form,
\begin{equation}
  |\Psi(0)\rangle = 
  \sum_{k_1,...,k_N}\psi(k_1 ,...,k_N)
  c_{k_1}^\dagger...c_{k_N}^\dagger |0\rangle,
\label{eqn:Psi0}
\end{equation}
which is the most general for fixed particle number.

The usual way to introduce a static vector potential $A {\hat{x}}$ is
to multiply the hopping parameters with a phase factor.  In this case
the $k$ vectors are shifted as $k \rightarrow k + \Phi$ with $\Phi= A
/\hbar c$, leading to
\begin{equation}
  \hat{H}(\Phi) = \sum_k \epsilon_{k+\Phi} \hat{n}_k + \hat{H}_i.
  \label{eqn:shiftedH}
\end{equation}
To arrive at Eq. (\ref{eqn:shiftedH}) one can also use the total
momentum shift operator on the total Hamiltonian constructed from
$\hat{H}$, and shift indices as was done to obtain
Eq. (\ref{eqn:Htg}).  In the same way one can obtain the wavefunction
corresponding to the shifted $\hat{H}(\Phi)$,
\begin{equation}
  |\Psi(\Phi)\rangle = 
  \sum_{k_1,...,k_N}\psi(k_1 +\Phi,...,k_N+\Phi)
  c_{k_1}^\dagger...c_{k_N}^\dagger |0\rangle.
\label{eqn:Psif}
\end{equation}

The criterion for the DC conductivity and the superfluid weight can
both be written~\cite{Shastry90,Scalapino92,Scalapino93} in the form
\begin{equation}
D = \frac{1}{2L} \frac{d^2 E(0)}{d \Phi^2}.
\label{eqn:d2Ddphi2}
\end{equation}
While the Drude weight and the superfluid weight quantities correspond
to different perturbations, the expression for these quantities
coincides, since in the above expression $\Phi=0$, hence the explicit
dependence on the vector potential, which gives rise to the
distinction, is neglected.  Taking advantage of the Hellmann-Feynman
theorem $D$ can be expressed as
\begin{equation}
  D = \frac{1}{2L}    \left\{
    \langle \Psi(0)| \frac{\partial^2 H(0)}{ \partial \Phi^2}|
    \Psi(0) \rangle  +
    \langle \frac{\partial \Psi(0)}{ \partial \Phi}| \frac{\partial
      H(0)}{ \partial \Phi}|    \Psi(0) \rangle   +
    \langle \Psi(0)| \frac{\partial H(0)}{ \partial \Phi}|    
\frac{\partial \Psi(0)}{ \partial \Phi} \rangle \right\} .
\label{eqn:Dck}
\end{equation}
The reason that both the Drude and superfluid weights can be written
in this form is due to the fact that Eq. (\ref{eqn:Dck}) is a linear
response expression in which the effect of the perturbing field is set
to zero.  

The derivatives with respect to $\Phi$ of the Hamiltonian can be made
to correspond with derivatives with respect to the momenta, i.e. it
holds that,
\begin{equation}
\label{eqn:dhdphi}
  \frac{\partial\hat{H}(\Phi)}{\partial \Phi} =
  \sum_k \frac{\partial\epsilon_{k+\Phi}}{\partial k} \hat{n}_k ,
\end{equation}
and
\begin{equation}
\label{eqn:d2hdphi2}
  \frac{\partial^2\hat{H}(\Phi)}{\partial \Phi^2} =
  \sum_k \frac{\partial^2\epsilon_{k+\Phi}}{\partial k^2} \hat{n}_k,
\end{equation}
\begin{equation}
\frac{\partial |\Psi(\Phi)\rangle }{\partial \Phi}  
 = 
  \sum_{k_1,...,k_N} \sum_i
\frac{ \partial \psi(k_1 +\Phi,...,k_N+\Phi)}{\partial k_i}
  c_{k_1}^\dagger...c_{k_N}^\dagger |0\rangle.
\label{eqn:dpsidphi}
\end{equation}
The derivative with respect to $k$ is
ambiguous~\cite{Scalapino92,Scalapino93}.  For a finite system with
size $L$ the summation in Eqs. (\ref{eqn:dhdphi}),
(\ref{eqn:d2hdphi2}), and (\ref{eqn:dpsidphi}) is defined on grid
points separated by $2\pi/L$ in reciprocal space.  Thus one way to
define the derivatives is using these grid points (for example the
finite element definition).    

The total momentum shift extended to length $nL$ extends the Hilbert
space, hence the derivatives can also be defined using the extended
states on the finer grid $2\pi/(nL)$.  Note that the summations in
Eqs. (\ref{eqn:dhdphi}), (\ref{eqn:d2hdphi2}), and
(\ref{eqn:dpsidphi}) are still defined on the grid $2\pi/L$.  When the
thermodynamic limit is taken $\epsilon_k$ is a continuous function,
hence this distinction between grids causes no ambiguity in the
application of Eqs. (\ref{eqn:dhdphi}) and (\ref{eqn:d2hdphi2}).  The
wavefunction, however, can be discontinuous, and, as discussed below,
this leads to consequences.  Using Eqs. (\ref{eqn:dhdphi}),
(\ref{eqn:d2hdphi2}), and (\ref{eqn:dpsidphi}) one can show that
\begin{equation}
  D = \frac{1}{2L}
\sum_{k} \left( \frac{\partial^2 \epsilon_k}{\partial k^2} n_{k}
+ 
\frac{\partial \epsilon_k}{\partial k} 
\left(
\frac{\partial n_{k}}{\partial k}
+ 
\sum_{k'} 
\frac{\partial n_{k,k'}^{(2)}}{\partial k'}
\right)\right), 
\label{eqn:Ds1s}
\end{equation}
where $n_k$ and $n_{k,k'}^{(2)}$ denote the one and two-body momentum
densities in the ground state, defined as
\begin{equation}
n_k = \sum_i \sum_{\stackrel{k_1,...,k_N}{k_i = k}}|\psi(k_1,...,k_N)|^2
\end{equation}
and
\begin{equation}
n_{k,k'}^{(2)} = \sum_{i\neq j} \sum_{\stackrel{k_1,...,k_N}{k_i = k,k_j =
k'}}|\psi(k_1,...,k_N)|^2.
\end{equation}
Eq. (\ref{eqn:Ds1s}) is arrived at by using Eqs. (\ref{eqn:dhdphi}),
(\ref{eqn:d2hdphi2}), and (\ref{eqn:dpsidphi}), and the identity
\begin{equation}
\langle 0 |\hat{c}_{k_N}...\hat{c}_{k_1} \hat{n}_k
\hat{c}_{k_1}^\dagger...\hat{c}_{k_N}^\dagger |0\rangle =
\sum_i \delta_{k_i k}.
\end{equation}

For the case $n=1$, we replace the derivative in Eq. (\ref{eqn:Ds1s})
by
\begin{equation}
  \frac{\partial n_k}{ \partial k} \rightarrow \frac{n_{k + 2\pi/L} -
n_k}{2\pi/L}.
\label{eqn:derivn1}
\end{equation}
This definition corresponds to the ``envelope function'' definition of
SWZ~\cite{Scalapino92,Scalapino93}.  To see this consider the system
at $\Phi=0$ and $\Phi=2\pi/L$.  The ground state at $\Phi=0$ is of the
form in Eq. (\ref{eqn:Psi0}), at $\Phi=2\pi/L$ it is
Eq. (\ref{eqn:Psif}), no longer the ground state in general.  For both
$\Phi=0$ and $\Phi=2\pi/L$ the ground state density is given by $n_k$.
In Eq. (\ref{eqn:derivn1}) the function $n_k$ (corresponding to the
ground state) is used in both cases.  When the thermodynamic limit
($L\rightarrow\infty$) is taken the first two terms in
Eq. (\ref{eqn:Ds1s}) cancel due to partial integration resulting in
\begin{equation}
  D^{(n=1)} = \frac{L}{8\pi^2}
\int \mbox{d}k\mbox{d}k'
\frac{\partial \epsilon_k}{\partial k} 
\frac{\partial n_{k,k'}^{(2)}}{\partial k'}.
\label{eqn:Ds}
\end{equation}
This quantity integrates to zero, due to the periodicity of the
Brillouin zone, unless, as discussed below, pairing occurs in the
two-body density.  These arguments allow association of $D^{(n=1)}$ 
with the superfluid weight.

We now consider the implications of the different properties of the
derivatives for $n=1$ and $n\rightarrow\infty$.  For segments for
which $n_k$ and $n_{k,k'}^{(2)}$ are continuous the two definitions of
the derivatives (based on the spacing $2\pi/L$ vs. $2\pi/(nL)$)
coincide, however this is not true when either densities are
discontinuous in $k$.  While on the larger grid $2\pi/L$ a
discontinuity in these quantities leads to a divergence, on the grid
$2\pi/(nL)$ the discontinuity does not occur when the derivative at
the $k$-grid points is evaluated and the limit $n\rightarrow\infty$ is
taken first, and the derivative is defined as adiabatically shifted.

As an example one can consider a Fermi sea, for which the term
depending on the two-body density does not contribute since there are
no correlations between momenta.  When a phase is applied the energy
levels and the momentum densities are shifted as $\epsilon_k
\rightarrow \epsilon_{k+\Phi}$, $n_k \rightarrow n_{k+\Phi}$.  If the
phase $\Phi\approx 2\pi/L$, and the ground state of the new
Hamiltonian is used in defining the derivative (``envelope
function''), then the discontinuity {\it contributes} to the
derivative, since if $n_k$ is the last filled state near the
discontinuity, then $n_{k+2\pi/L}$ will be the first unfilled one.
However, for small $\Phi$ (which corresponds to the limit
$n\rightarrow\infty$) if $n_k$ corresponds to the last filled state
then $n_{k+\Phi}$ does not change.  Excluding the discontinuities
(which are relevant to the second term in Eq. (\ref{eqn:Ds1s})) from the
partial integral leads to
\begin{equation}
  D^{(n\rightarrow\infty)} = \frac{1}{2\pi} \Delta n_{k_F} 
  \frac{\partial \epsilon_{k_F}}{\partial k},
\label{eqn:Dckfs}
\end{equation}
where the discontinuities are assumed to be at $k=\pm k_F$ (Fermi wave
vector).  When spin is included then each spin component will
contribute a term of the form in Eq. (\ref{eqn:Dckfs}).  For this
reason we associate the quantity $D^{(n\rightarrow\infty)}$ with the
Drude weight.

To explore the connection between conduction and the discontinuity in
the momentum density further we consider the quantity
\begin{equation}
\Pi(y) = \left| \langle \Psi |\hat{U}(y)| \Psi \rangle \right| =
\left|\sum_{k_1,...,k_N}\psi^*\left(k_1+y,...,k_N+y
\right) \psi(k_1,...,k_N)\right|.
\label{eqn:psiupsi}
\end{equation}
The quantity $(L^2/(2\pi^2)\mbox{Re}
\hspace{.2cm}\mbox{ln}\Pi(2\pi/L)$ was suggested by Resta and Sorella
as a criterion of localization.  As a result of Kohn's
hypothesis~\cite{Kohn64} localization is also a criterion to
distinguish conductors from insulators.  If the wavefunction
$\psi(k_1,...,k_N)$ is a continuous functions of its arguments then
$\Pi(2\pi/L)$ approaches unity in the limit of large system size.  The
functions $n_k$ and $\Pi(y)$ are then continuous, corresponding to
insulation.  When $n_k$ is discontinuous then the magnitude of the
wavefunction $\psi(k_1,...,k_N)$ is also discontinuous.  In the
following we assume that the magnitude of $\psi(k_1,...,k_N)$ is
discontinuous but its phase is not.  Since $\psi(k_1,...,k_N)$
describes indistinguishable particles, the discontinuity has to occur
as a function of any of its arguments.  Moreover, on physical grounds
we anticipate that this discontinuity occurs at the Fermi wave-vector.
The effect of the discontinuity can be assessed by considering the
difference
\begin{eqnarray}
\nonumber
\Pi(0) - \Pi(\epsilon) = & 
\left|
\frac{1}{(2\pi)^N}\int \mbox{d}k_1.\mbox{d}k_N
\psi^*(k_1,...,k_N)\psi(k_1,...,k_N)
\right| \\
& -
\left|
\frac{1}{(2\pi)^N}\int \mbox{d}k_1.\mbox{d}k_N
\psi^*(k_1+\epsilon,...,k_N+\epsilon)
 \psi(k_1,...,k_N)
\right|.
\label{eqn:1mu}
\end{eqnarray}
where $\epsilon$ denotes an infinitesimal and the thermodynamic limit
was taken.  The integrands in the first term and the second term will
cancel for regions where the coefficient $\psi(k_1,...,k_N)$ is
continuous.  The contribution of a discontinuity at $k_F$ will be of
the form
\begin{equation}
\rho(k_F+;k_F+) + \rho(k_F-;k_F-) - \rho(k_F+;k_F-) - \rho(k_F-;k_F+),
\end{equation}
where $\rho(k;k')$ denotes the one-body density matrix in $k$-space.
Rewriting in a natural orbital representation this contribution takes the form
\begin{equation}\sum_i q_i [
\gamma_i^*(k_F+) \gamma_i(k_F+) +
\gamma_i^*(k_F-) \gamma_i(k_F-) -
\gamma_i^*(k_F+) \gamma_i(k_F-) -
\gamma_i^*(k_F-) \gamma_i(k_F+)],
\end{equation}
(with $0\leq q_i \leq 1$ and $\gamma_i(k)$ denoting the natural orbitals)
which under the assumption of a continuous phase is a positive
quantity.  Since this is also the case for $\Pi(0)-\Pi(-\epsilon)$ it
follows that for discontinuous coefficient $\psi(k_1,...,k_N)$ the
function $\Pi(y)$ will contain a $\delta$-function contribution at the
origin.  These results coincide exactly with the results of Resta and
Sorella~\cite{Resta99} where a function of the quantity $|\Pi(2\pi/L)|$
is suggested as a criterion of localization and conduction.

To understand the effect of pairing we study the BCS wavefunction 
\begin{equation}
|\Psi_{BCS}\rangle = \prod_k (u_k + v_k c_{k\uparrow}^\dagger 
c_{-k\downarrow}^\dagger) |0\rangle.
\end{equation}
We assume a BCS Hamiltonian with constant coupling between Cooper
pairs.  Calculating the properties of this wavefunction requires
generalization to include spin and variable particle number of
Eq. (\ref{eqn:Ds1s}) which presents no difficulty.  Since the one-body
density of the BCS wavefunction is continuous the first two terms
cancel by partial integration when the thermodynamic limit is taken.
Thus we are lead to consider the last term only, which depends on the
two-body momentum density $n_{k,k'}^{(2)}$.  This quantity can be
broken up into components with parallel and anti-parallel spins.  The
parallel spin two-body density is again continuous, hence does not
contribute.  The two-body density when the spins are anti-parallel
gives
\begin{equation}
n_{k,k'}^{(2)} = \left\{ \begin{array}{rl}
  f(k') & k' = -k \\
  f(k)f(k')   & k' \neq -k,
       \end{array}  \right.
\end{equation}
with
\begin{equation}
f(k) = \frac{|v_{k}|^2}{|u_{k}|^2 + |v_{k}|^2}.
\end{equation}
Explicit calculation for the BCS wavefunction then yields for the
thermodynamic limit
\begin{eqnarray}
\nonumber
D = \frac{1}{4\pi}\sum_{\sigma} \int \mbox{d}k 
\left(-\frac{\partial \epsilon_k }{\partial k}
\frac{\partial n_{k\sigma} }{\partial k}\right) 
+ \frac{L}{8\pi^2}
\sum_{\sigma} 
\int \mbox{d}k\mbox{d}k' 
\frac{\partial \epsilon_k }{\partial k} n_{k\sigma}
\frac{\partial n_{k'-\sigma} }{\partial k}.
\end{eqnarray}
The first term arises since $n^{(2)}_{k\sigma,-k-\sigma}=n_{k\sigma}$,
i.e. due to Cooper pairing.  Due to the continuity of $n_{k\sigma}$ the last
term is zero.  Partial integration then results in
\begin{equation}
D = \frac{1}{4\pi} \sum_{\sigma}  \int \mbox{d}k 
\frac{\partial^2 \epsilon_k }{\partial k^2} n_{k\sigma} .
\end{equation}
Since the function $f(k)$ is continuous this result holds for both
$n=1$ and $n\rightarrow \infty$.  The result that the second
derivative of the ``envelope'' function of the ground state energy is
finite for a superfluid and zero for a normal metal was obtained for
the case of a ring with finite thickness by Byers and
Yang~\cite{Byers61}.

SWZ have also shown~\cite{Scalapino93} that for dimensions higher than
one the first non-adiabatic crossing occurs at zero field when the
thermodynamic limit is taken.  This leads to a distinction between
evaluating $\frac{\partial^2 E(\Phi)}{\partial \Phi^2}$ first and then
taking the thermodynamic limit or vice versa.  In generalizing the
formalism presented here to higher dimensions one has to consider that
the differential operators in the superfluid and Drude weights operate
in one particular direction (that of the perturbing field).  If the
thermodynamic limit is first taken in the direction perpendicular to
the perturbing field, then the discontinuity can ``disappear''.  For
example, a two-dimensional non-interacting system at half-filling has
a discontinuous momentum density, $n_{k_x,k_y}$, but the function
$f(k_x) = \int \mbox{d}k_y n_{k_x,k_y}$ is a continuous function.
However, the definition of the derivative corresponding to the case
$n\rightarrow \infty$ resolves this ambiguity.  In that case
irrespective of the order of limits the discontinuity will be excluded
from the integration, as argued above for the Fermi sea.  Moreover, as
shown above, the discontinuity in the momentum density contains
exactly the same information as the localization order parameter of
Resta and Sorella~\cite{Resta99}, a quantity which is also insensitive
to dimensionality.

In conclusion the second derivative of the ground state energy with
respect to a perturbing field (vector potential) at zero field was
derived and shown to be an expectation value over the one and two-body
momentum densities.  A length scale associated with the perturbation
was defined, and through it states with twisted boundary conditions
were introduduced, allowing for the possibility of defining the
adiabatic derivative (Drude weight) and the derivative of the ground
state energy envelope function (superfluid weight).  The resulting
expression for the Drude weight is not the zero frequency limit of an
quantity based on time-dependent perturbation theory.  The Drude
weight is finite in the presence of discontinuities in the
wavefunction (which correspond to discontinuities in the momentum
densities), as well as due to BCS pairing.  The superfluid weight is
not sensitive to discontinuities in the momentum densities, but is
finite in the presence of BCS pairing.  It was shown that a
localization quantity suggested by Resta and Sorella~\cite{Resta99}
based on a tenet of Kohn~\cite{Kohn64} contains the same information
as the discontinuity in the momentum density.  Thus the connection
between the localization hypothesis of Kohn~\cite{Kohn64} and the
criterion of metallicity in the Landau theory of Fermi liquids is
established.

\section*{Acknowledgements}The author is indebted to Hans Gerd Evertz
for helpful discussions.  Part of this work was performed at the
Institut f\"ur Theoretische Physik at TU-Graz under FWF grant number
P21240-N16.

\end{document}